\documentclass[11pt, singlecolumn, logo, copyright]{googledeepmind}

\usepackage[authoryear, compress, round]{natbib}
\usepackage{hyperref}
\usepackage{url}
\usepackage{placeins}
\usepackage{float}
\usepackage{graphicx}
\usepackage{subcaption}
\usepackage{listings}
\usepackage{xcolor}

\title{Representation biases: will we achieve complete understanding by analyzing representations?}

\author[1]{Andrew Kyle Lampinen}
\author[1]{Stephanie C. Y. Chan}
\author[1]{Yuxuan Li}
\author[1]{Katherine Hermann}

\affil[1]{Google DeepMind}

\correspondingauthor{lampinen@google.com}


\reportnumber{} 

\renewcommand{\today}{}

\begin{abstract}
A common approach in neuroscience is to study neural representations as a means to understand a system---increasingly, by relating the neural representations to the internal representations learned by computational models. However, a recent work in machine learning \citep{lampinen2024learned} shows that learned feature representations may be biased to over-represent certain features, and represent others more weakly and less-consistently. For example, simple (linear) features may be more strongly and more consistently represented than complex (highly nonlinear) features. These biases could pose challenges for achieving full understanding of a system through representational analysis. In this perspective, we illustrate these challenges---showing how feature representation biases can lead to strongly biased inferences from common analyses like PCA, regression, and RSA. We also present homomorphic encryption as a simple case study of the potential for strong dissociation between patterns of representation and computation. We discuss the implications of these results for representational comparisons between systems, and for neuroscience more generally. 
\end{abstract}

\begin{document}

\maketitle

\begin{figure}[htp]
\centering
\includegraphics[width=0.91\linewidth]{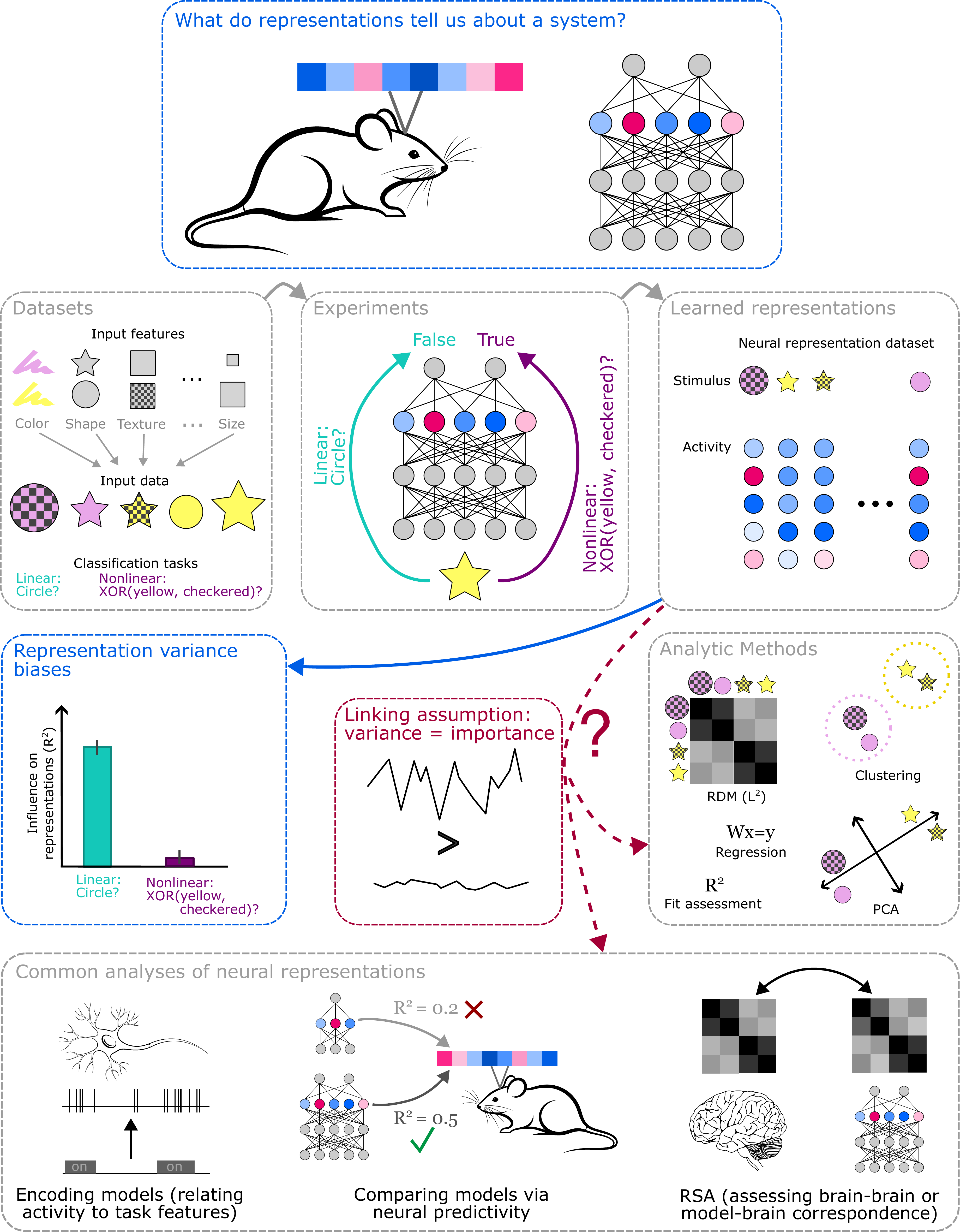}
\caption{What do representations tell us about a system? Here, we discuss some experiments by \citet{lampinen2024learned} that illustrate challenges in making inferences about a system from its representations. The experiments involve training models on controlled datasets mapping from many input features to outputs involving simpler or more complex functions of the input features. The representations that models learn from these datasets show substantial biases towards certain outputs over others (e.g. towards simple linear functions of the input over complex nonlinear ones), even when the model has reliably learned both. These biases raise questions about a linking assumption that underlies many analytic methods and common analyses in neuroscience: that variance explained in the neural representations (or activity) is a good proxy for the importance of a signal.} \label{fig:overview}
\end{figure}

\section{Introduction}

A central approach of neuroscience is analyzing patterns of neural representation to learn about a system \citep{kriegeskorte2019peeling}. In particular, computational neuroscience has increasingly relied on relating patterns of neural activity to the internal representational structures of computational models \citep{churchland1990neural,kriegeskorte2008representational,sucholutsky2023getting,feather2025brain}. 
However, there are philosophical questions about how to justify interpreting internal activity as representations \citep{shea2018representation,fallon2023we,cao2022putting}
and conceptual and practical challenges to understanding a system through analyzing its internal activity or representations \citep{poldrack2006can,marom2009precarious,jonas2017could,ritchie2019decoding,sexton2022reassessing,dujmovic2024inferring}.

Here, we use case studies from recent work in machine learning to illustrate some practical challenges to understanding a system's function by studying its internal representations (see Fig. \ref{fig:overview} for an overview)---and discuss the implications for neuroscience.
In particular, we focus on the results of \citet{lampinen2024learned}. In this work, the authors study the relationship between patterns of representation and computation in machine learning models, using controlled experiments. The authors identify substantial biases in the learned representations: some features are much more strongly represented than others, even if they play similar computational roles in the system's behavior. These representation biases mean that common analytic methods---such as Principal Component Analysis (PCA), Representational Similarity Analysis \citep[RSA;][]{kriegeskorte2008representational}, and linear regression---may be biased towards capturing some computational features over others. Thus, the many types of neuroscience experiments that use these methods like these to study a system's representations or activity may provide a biased picture of its computations.

We first briefly review the prior literature on representations and computations in cognitive science and neuroscience. We then turn to our central examples illustrating representation biases, and how they impact commmon representational analyses. We then briefly present a case study of homomorphic encryption, which illustrates how strong the dissociation between representations and the system's function can be. We close by discussing the implications of our results, and the questions they raise.

\subsection{Background}


In a foundational paper, \citet{marr2010vision} famously posited three levels of analysis for understanding an information processing system---computational, algorithmic, and implementational. In the years since, there has been substantial discussion about the (non)independence of the levels, their interactions and joint constraints, and what they imply about representation \citep[e.g.][]{mcclamrock1991marr,churchland1990neural,obrien2009role,stevens2012vision,piccinini2014foundations,hardcastle2015marr}.
In the philosophical literature, similarly, the question of the relationship between representation and computation has been somewhat fraught \citep[e.g.][]{eliasmith2003neural,piccinini2008computation,vangelder2014compositionality,shea2018representation}.  

There have also been arguments from a variety of perspectives that intelligence is best understood (or engineered) by focusing on the interaction between a system and its environment  \citep{clark1998being,cisek1999beyond,brooks1991intelligence,varela2017embodied}. These approaches therefore tend to think of ``representations'' as more dynamic and entangled with the environment state rather than as static ways of encoding information.

In addition to the conceptual debates about representation, there has been substantial debate about the empirical relationship between the representations of a system and its computational mechanisms or behavior. 
Several works have highlighted particular case studies where applying neuroscience analyses to well-understood computational systems does not yield the expected results \citep{marom2009precarious,jonas2017could}.
In machine learning, aside from the main paper we discuss, a number of other works have demonstrated that there can be substantial dissociations between the internal representational structure of a system and its computations or behavior \citep{hermann2020shapes,braun2025not}---especially under distribution shift \citep{friedman2023comparing,dujmovic2024inferring}.

There has also been a longer-standing discussion about the more concrete issue of the relationship between neural activity and cognitive mechanisms.
For example, many researchers have noted that making inferences about cognitive processes or information content from downstream effects like fMRI activity \citep{poldrack2006can} or decoding \citep{ritchie2019decoding} can be logically invalid. Others have noted the possibility that representational correlations do not necessarily imply computational correspondence, and therefore emphasized more stringent measures \citep{sexton2022reassessing}. As in the conceptual work above, some neuroscientists have taken more provocative stances that ``representation'' is generally the wrong way to understand neural activity \citep[e.g.][]{freeman1990representations}.

However, there is clearly still information to be gained from representational analyses, and thus many researchers take a \emph{pragmatic} perspective on representation in neuroscience: interpreting activity measures as a representation insofar as it satisfies conditions of reliable decodability together with causal efficacy (\citealp{cao2022putting}; cf. \citealp{poldrack2021physics,baker2022three}). That is, a pattern (or distribution) of activity can be considered a representation insofar as we can consistently decode the expected information content from it, and intervening on it changes the system's behavior in the expected way---e.g. if we can identify a consistent pattern of neural activity corresponding of a stimulus, and reactivating that representation produces the stimulus response even when the stimulus is absent. We adopt a similar perspective here (and the ``representations'' studied here satisfy both criteria).

While the issues we focus on here are related to some of these overlapping discussions, we focus more on the question of \emph{how} we identify candidate feature representations in the first place---and whether the types of feature representations that we identify might be systematically biased towards those that serve certain kinds of computations.

\section{Representation biases}
\begin{figure}[htp]
\centering
\begin{subfigure}[b]{0.22\textwidth}
\centering
\captionsetup{width=.95\linewidth}
\includegraphics[width=\linewidth]{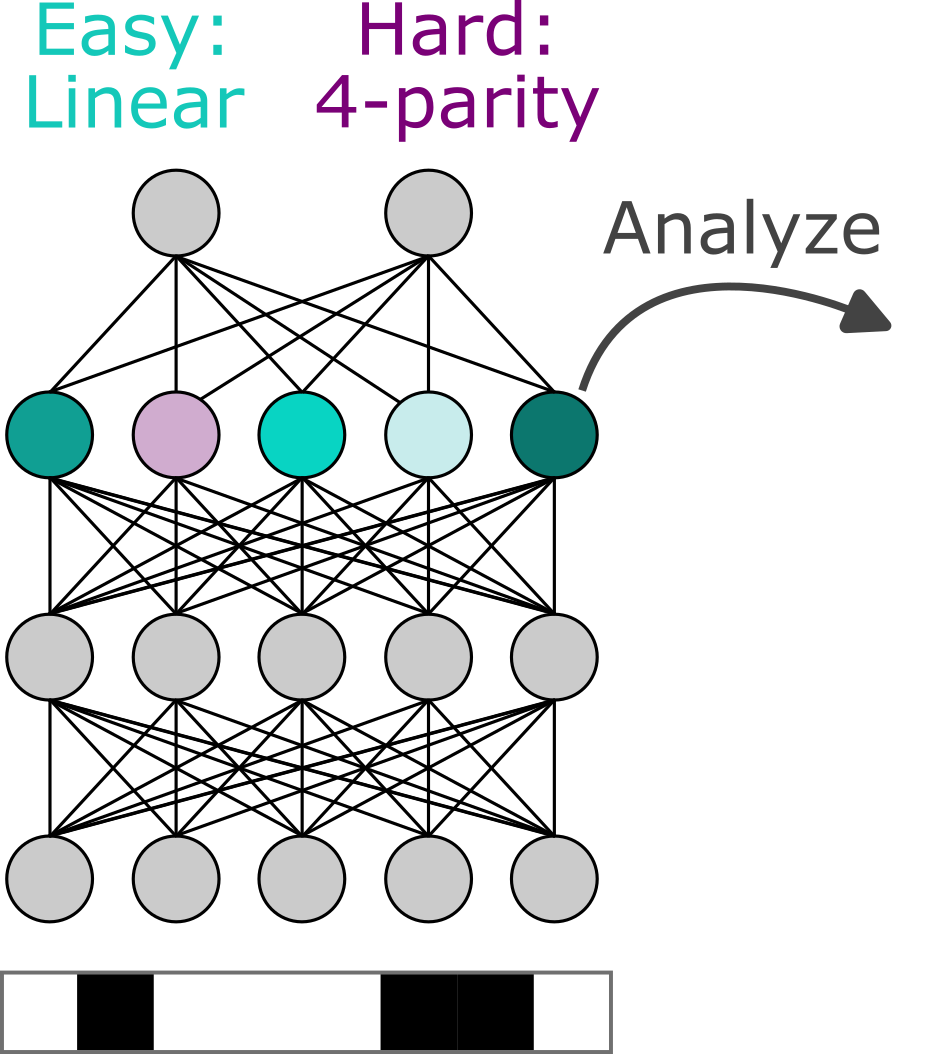}
\caption{Model\\representations.}\label{fig:biases_pca:model}
\end{subfigure}
\begin{subfigure}[b]{0.18\textwidth}
\centering
\includegraphics[width=\linewidth]{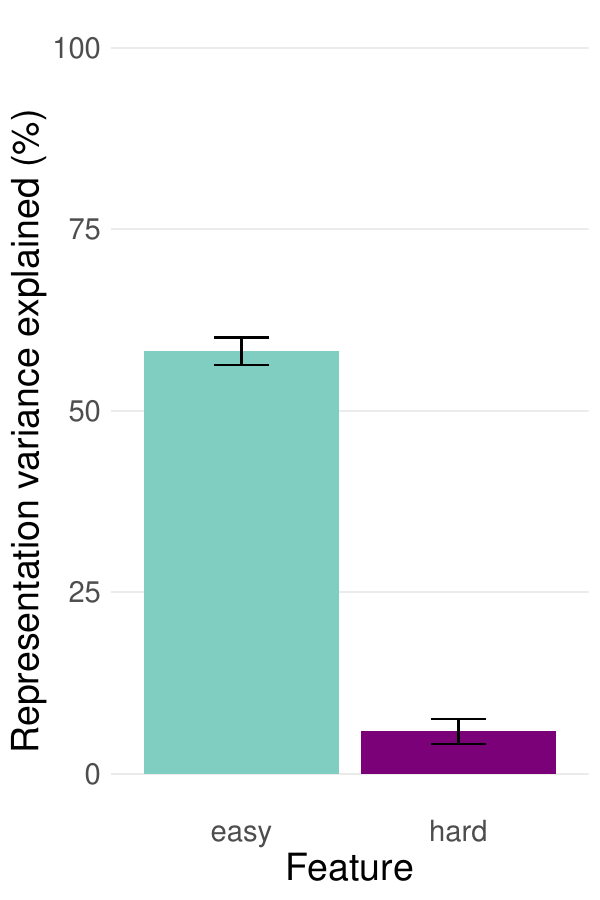}
\caption{Representation variance biases.} \label{fig:biases_pca:var_bias}
\end{subfigure}%
\begin{subfigure}[b]{0.33\textwidth}
\centering
\captionsetup{width=.8\linewidth}
\includegraphics[width=\linewidth]{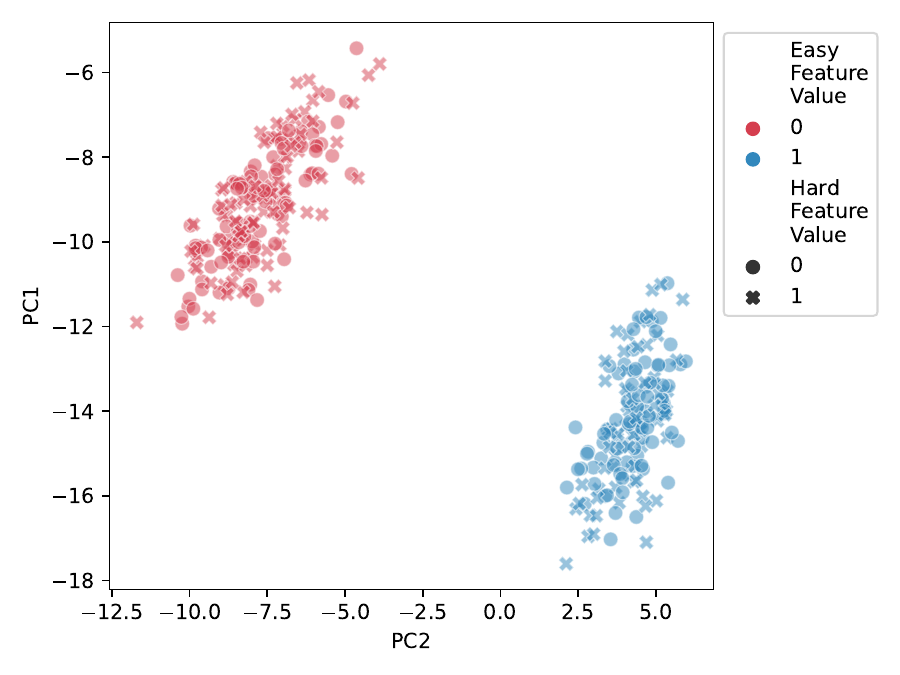}
\caption{Top principal components of representations.}\label{fig:biases_pca:pca}
\end{subfigure}%
\begin{subfigure}[b]{0.25\textwidth}
\centering
\captionsetup{width=.8\linewidth}
\includegraphics[width=\linewidth]{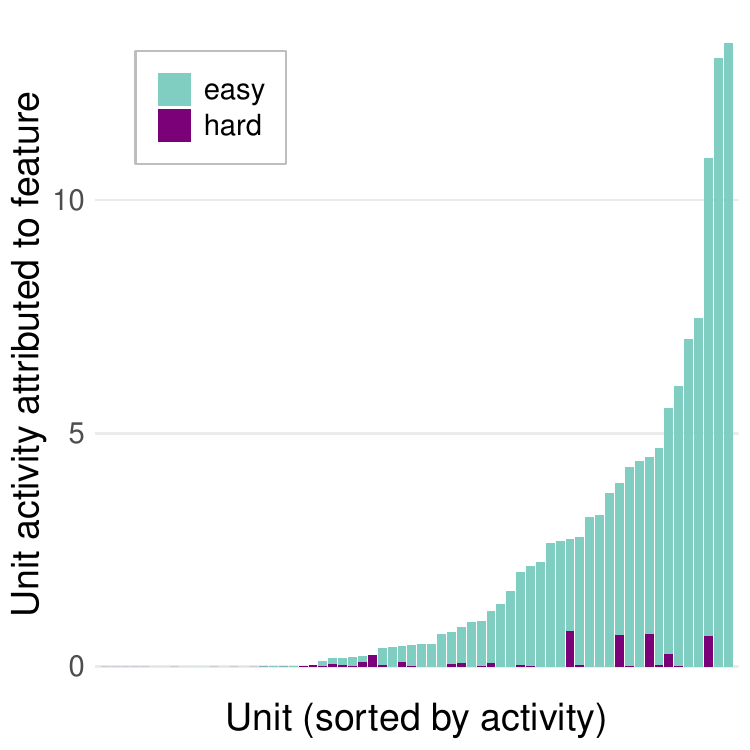}
\caption{Unit-level biases.\\}\label{fig:biases_pca:units}
\end{subfigure}%

\caption{Model representations can be substantially biased towards certain features, in the sense that those features carry much more variance in the representations than others that have similar computational roles. These analyses focus on the representations at the penultimate layer of an MLP model (\subref{fig:biases_pca:model}) trained to independently classify two features of a binary-valued input vector: one easy linear feature, and one hard highly-nonlinear feature. Although the final-layer task of the model is the same for both features---binary classification from the penultimate representation---the model's penultimate layer representations are heavily biased towards the easier feature (\subref{fig:biases_pca:var_bias}), with over 50\% of the variance in these penultimate representations explained by the easy feature, and only around 5\% by the hard feature. Thus, in the top principal components (\subref{fig:biases_pca:pca}) the representations are clearly clustered by the easy feature (colors), but the hard feature value (shapes) does not have a clear influence. At the individual unit level (\subref{fig:biases_pca:units}), the biases are equally clear---most of the units respond to the easy feature, and even those units that encode information about the hard feature tend to have more of their activity explained by the easy feature.} \label{fig:biases_pca}
\end{figure}

Here, we introduce the results of \citet{lampinen2024learned}. One kind of bias explored by the authors is depicted in Fig. \ref{fig:biases_pca}. A simple multi-layer perceptron model is trained to perform independent binary classification on an ``easy'' (linear) and ``hard'' (nonlinear) feature of binary input vectors. The model is trained on a sufficiently large dataset that it generalizes perfectly to a held-out test set for either feature; thus, the model is computing these two functions in a way that generalizes reliably. The representational analyses focus on the representations at the penultimate layer of the model, immediately before the final binary classification. Although the model is performing the same type of output classification on both features---i.e. a final linear classification---the \emph{variance explained} in the representations by the easy feature is an order of magnitude higher than that explained by the harder feature. 

Furthermore, this bias is also evident at the single-unit level. More units encode information about the easy feature, and even those that encode information about the harder feature still tend to have more of their variance explained by the easy feature---especially for the most active units.

Because of this variance bias, when analyzing this model's representations, many analyses tend to be biased towards the easy feature. For example, Principal Components Analysis (PCA) identifies precisely the dimensions in the representations that carry the most variance. Thus, the first several PCs show clear clusters according to the easy feature, and little organization according to the more difficult feature.

\begin{figure}[th]
\centering
\begin{subfigure}[t]{0.4\textwidth}
\includegraphics[width=\linewidth]{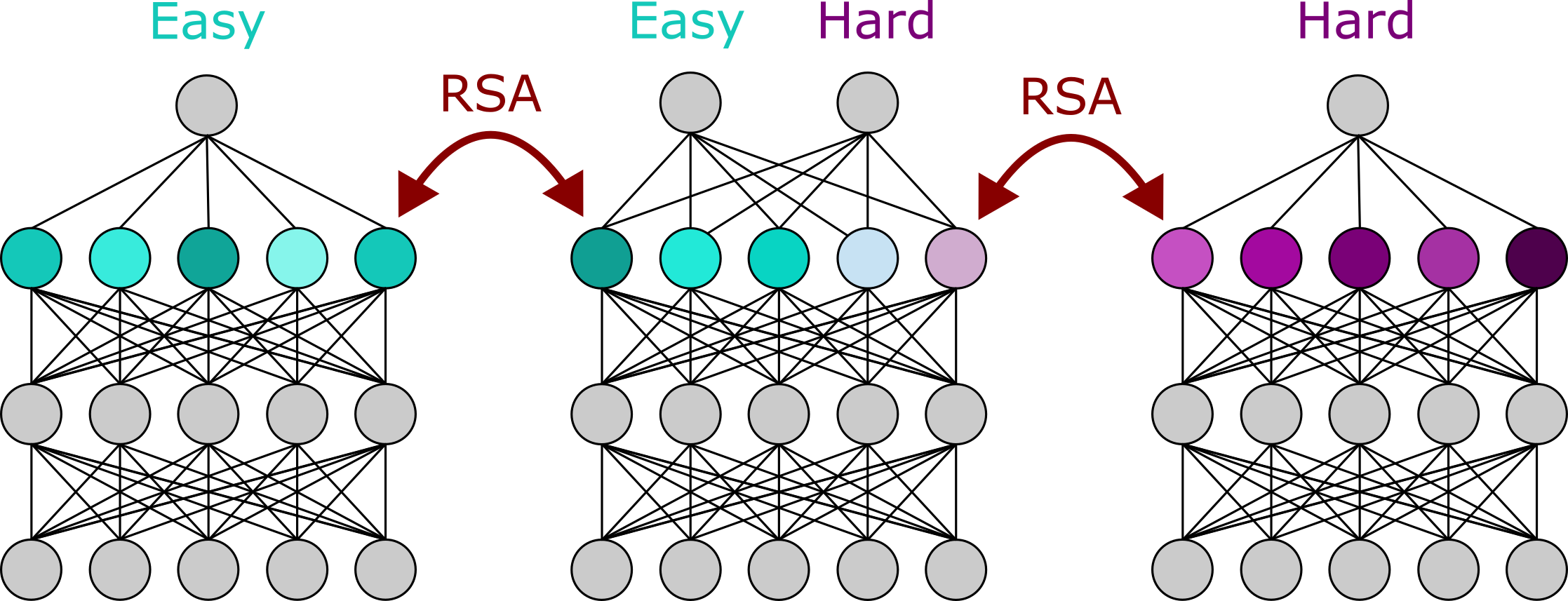}
\end{subfigure}
\begin{subfigure}{0.15\textwidth}
\phantom{blah}
\end{subfigure}
\begin{subfigure}[t]{0.4\textwidth}
\centering
\includegraphics[width=0.63\linewidth]{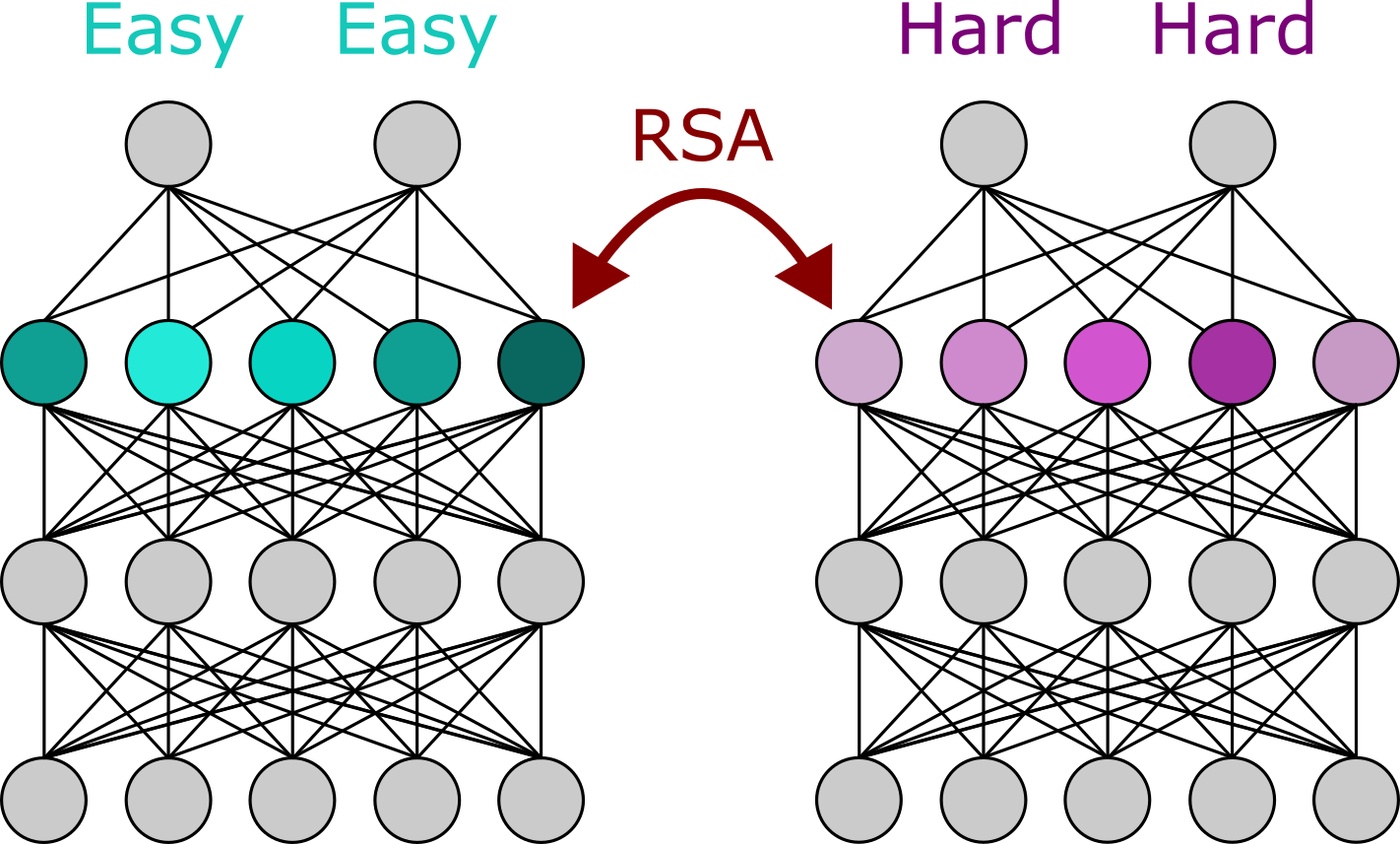}
\end{subfigure}\\[0.5em]
\begin{subfigure}[t]{0.4\textwidth}
\includegraphics[width=\linewidth]{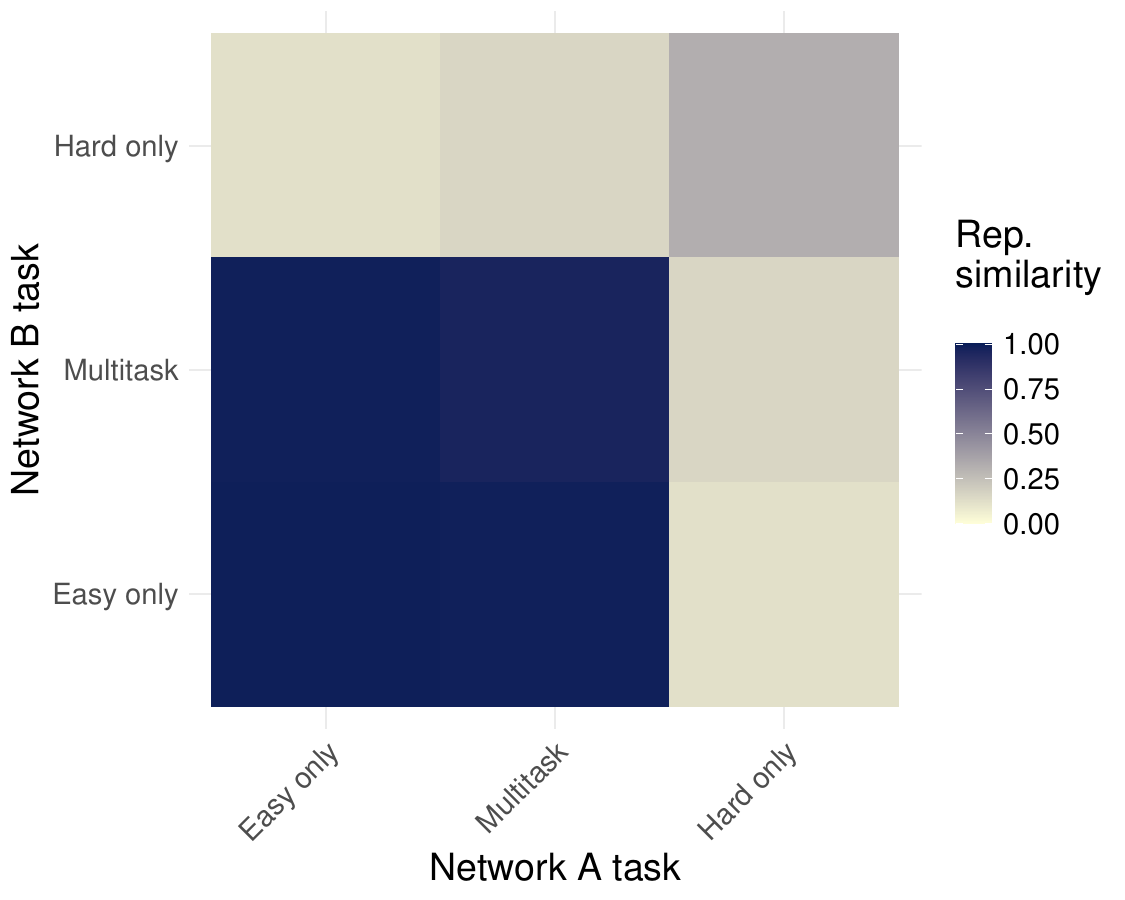}
\end{subfigure}
\begin{subfigure}{0.15\textwidth}
\phantom{blah}
\end{subfigure}
\begin{subfigure}[t]{0.4\textwidth}
\includegraphics[width=\linewidth]{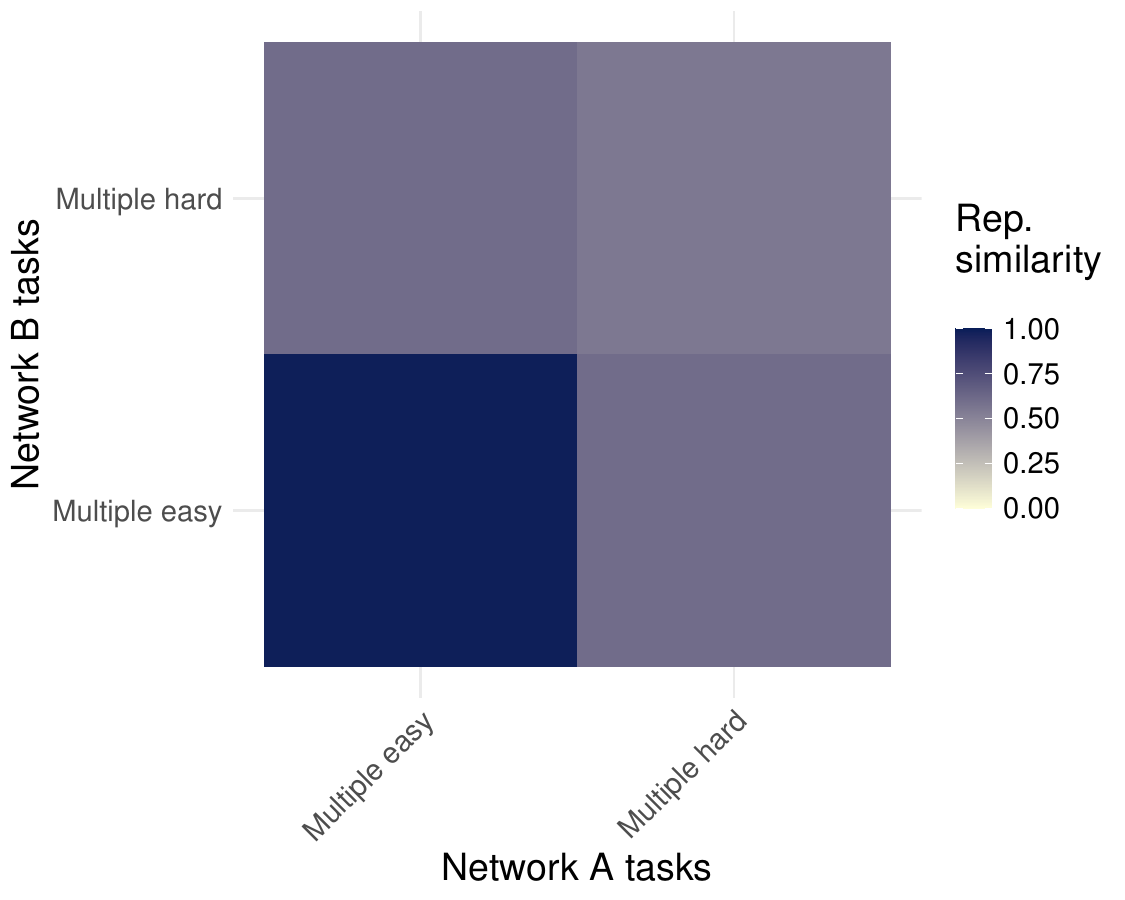}
\end{subfigure}%
\caption{Representation biases have downstream impacts on analyses that compare representations between systems. Here, we show that RSA within and between different sets of models can give surprising results due to representation biases. In each plot, ideally the blocks on the diagonal would be darkest (i.e., show the highest similarity), and the blocks off the diagonal would show graded similarity corresponding to the functional overlap. However, that is not the case. (Left) When comparing a model trained to output both easy and hard features to ones that are trained on only one feature, the multi-task model appears very similar to the easy-task only model \citep{hermann2020shapes}. In fact, the models trained only on the hard task do not even appear particularly similar to other models trained on the same exact task. (Right) When models are trained on multiple easy or multiple hard tasks, the models trained on only hard tasks appear \emph{less} similar to other models trained on exactly the same tasks than they do to models trained on strictly easier tasks that use the same input units. (5 models were trained for each combination of features.)} \label{fig:rsa}
\end{figure}

Beyond PCA, \emph{many} analyses treat variance as a proxy for the importance of a signal. Thus, these biases also affect other analyses like regression (onto the unit-level or population representations, e.g. from task features, a computational model, or another neural system) or RSA. Fig. \ref{fig:rsa} demonstrates how the biases can give surprising results in RSA. For example, when a model that is trained to compute both easy and hard features is compared to separate models that compute \emph{only} the easy feature, or \emph{only} the hard feature, the representations of the multi-task model appear very similar to those of the easy-task-only model, and very different from those of the hard-task-only model \citep[cf.][]{hermann2020shapes}. In some cases, models that are trained to compute only hard features can appear \emph{less} similar to other models trained on the same tasks than they do to models that are computing easy features from overlapping inputs. Thus, the models that have the most similar representations are not necessarily the ones that are computing the most similar functions. These results are supported by several other recent machine learning studies that similarly highlight the possibility for dissociations between representation and computation \citep{friedman2023comparing,braun2025not}.

\subsection{Why are representations biased?}

\begin{figure}[th]
\centering
\includegraphics[width=\linewidth]{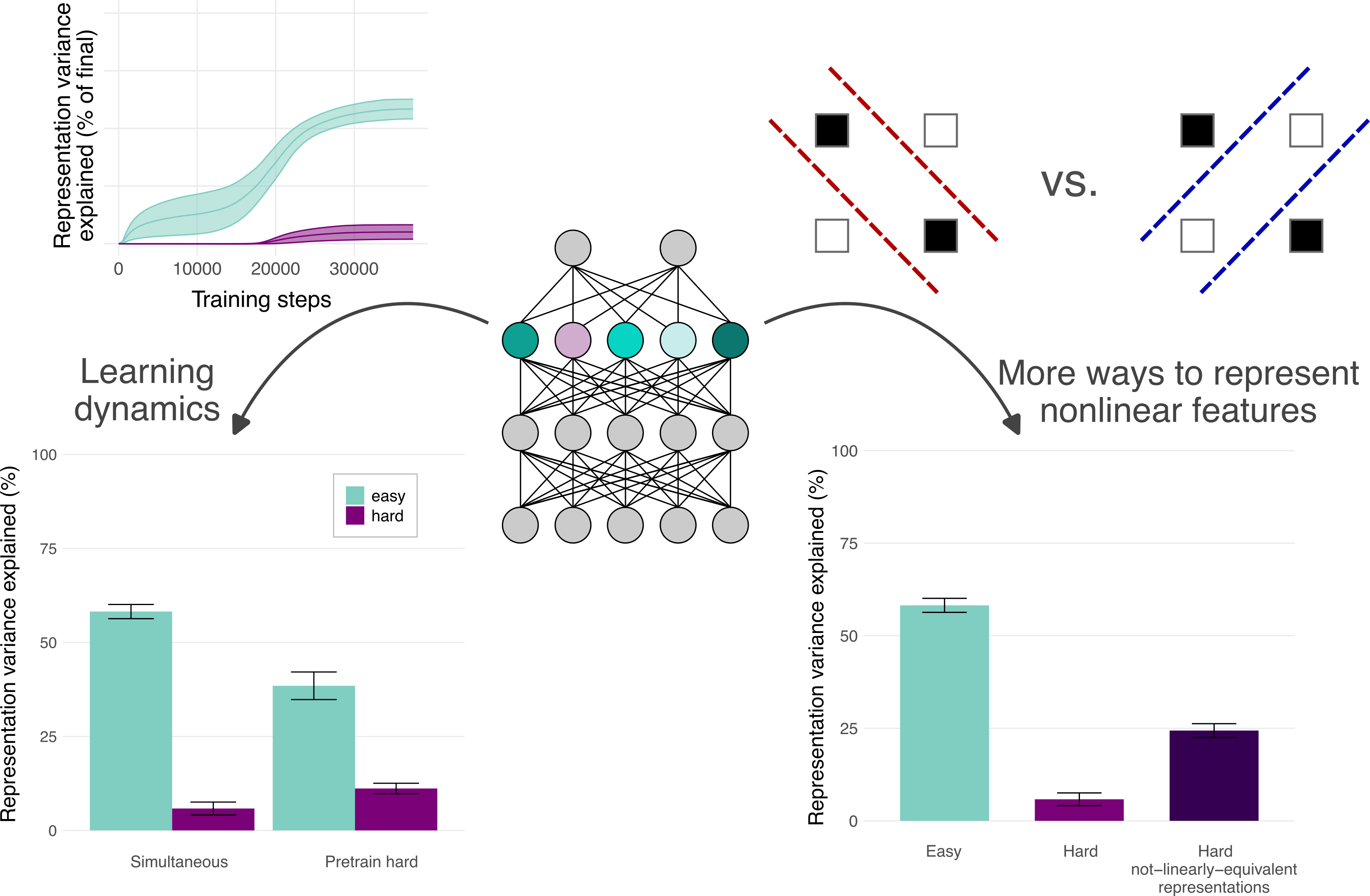}
\caption{Why are representations biased towards easier features? The biases are driven by multiple factors, including learning dynamics and the different ways that nonlinear features can be represented. (Left) By manipulating training order (training the hard task first rather than both simultaneously), the magnitude of the biases can be reduced. (Right) Likewise, by accounting for the fact that there can be more ways to represent a nonlinear feature that are not linearly equivalent---for example, different ways of drawing intermediate classification boundaries to compute an XOR function---we can identify other components of the representations that may be contributing to the model's computation of the hard feature. Together the learning dynamics and multiple ways of representing features explain most of the representation bias towards the easy feature over the hard.} \label{fig:bias_origins}
\end{figure}

\citet{lampinen2024learned} also investigate the origins of the above biases; their conclusions are illustrated in Fig. \ref{fig:bias_origins}. In short, multiple factors contribute to the biases. The most straightforward factor is that features that are learned first tend to be favored in the representations. Simpler features tend to be learned before more complex ones \citep[cf.][]{huh2021low,shah2020pitfalls}, and this bias has been suggested to be a contributor to the generalization of deep learning \citep[e.g.][]{huh2021low}. 
This pattern can also be seen as an instance of the ``availability'' biases highlighted by \citet{hermann2023foundations}. 
Indeed, \citet{lampinen2024learned} showed that other features that change which features are learned first (such as prevalence of a feature in the dataset) also drive similar representation biases. When learning order is reversed by pretraining the hard feature, the representational gap between the easy and hard features narrows (Fig. \ref{fig:bias_origins} bottom left). However, learning order does not fully explain the difference between the two features.

The other factor is that linear and nonlinear features yield different patterns of representation. All the \emph{natural}\footnote{E.g. the ways that a feature could be represented in the minimum-capacity two-layer network that computes it.} ways of representing a linear feature are equivalent up to linear transformation. By contrast, there can be more ways to represent a nonlinear feature that are not equivalent to one another. For example, to compute the XOR function, there are several ways of drawing the classification boundaries (Fig. \ref{fig:bias_origins} top right) that yield different representational similarity structures---these different modes of representation discard different pieces of information, so it is not possible to (linearly or nonlinearly) transform from one to the other with complete accuracy. To detect the effect of these different patterns of representation, the authors regress from an augmented space from which all representation patterns are linearly decodable. Indeed, accounting for multiple patterns of representation explains part of the gap between the linear and nonlinear features.
In high-dimensional representation spaces, these multiple not-linearly-equivalent solutions tend to dilute the variance due to a computation across more representational dimensions that vary in dissimilar ways, resulting in patterns of activity that are less similar between models trained on the same task, and less variance attributable directly to the nonlinear feature in question.

It is natural to ask whether the differences in representation patterns for the different types of features should really be interpreted as a ``bias''---after all, they do correspond to meaningful differences in computations (and how they map onto other levels of analysis) between the two features. We will return to this issue in the discussion. However, regardless of how they are interpreted, these differences still lead to biases in our inferences from representations, e.g. when using analyses like RSA or PCA they lead us to focus on representations used for computing only some of the features a system is using.

\textbf{The generality of feature biases:} While for brevity we focus on biases due to feature complexity here, \citet{lampinen2024learned} also find similar biases driven by other factors like prevalence in the data or position in a time sequence. Furthermore, while the examples here are taken from MLPs, there are similar types of biases in transformers and convolutional networks. Furthermore, the biases generally persist under various optimizers, regularization like dropout or weight decay, etc. Concurrent works have also observed similar biases in  larger models trained on more naturalistic data \citep{fel2024understanding}. Thus, these biases seem to be a fairly general phenomenon of feature learning. 

\section{How might feature biases impact neuroscience?}

Given that neuroscience often relies on analysis of representations, the kinds of biases discussed above could impact inferences in neuroscience as well.

For example, a growing area of research relies on comparing representations between AI models and humans or animals \citep{sucholutsky2023getting}, and there have been explicit calls for studying representational as well as behavioral alignment of models to the brain \citep[e.g.][]{feather2025brain}---as well as collaborative benchmarks for this purpose \citep[e.g][]{schrimpf2020integrative}. However, if the AI models are biased towards representing certain features more strongly, this could lead to biased inferences from these analyses and benchmarks. For example, such biases might help to explain why there is surprisingly little difference between different AI model architectures in how closely they align with brain representations \citep{conwell2023can}---if the representations of the various models tend to be dominated by the simplest (or most prevalent) features which any objective will learn. 

However, even for other types of analyses, the existence of representation biases raises questions. Do natural systems similarly produce representations that are biased by these (or other) factors? If so, then the kinds of computations and similarities that we tend to discover in these systems may likewise be biased. For example, we might overestimate how much high representational alignment between two humans (or animals) tells us about the alignment of their computations---the systems might have high representational alignment even if they only align on some features.

Conversely, it is common practice to compute a noise ceiling for comparisons by using between-individual predictions; e.g. using between-subject neural predictivity as an upper bound on how well a model could be expected to predict neural data. However, the fact that some features can be computed in ways that yield non-equivalent patterns of representation means that interpreting such measures as a ``noise'' ceiling could be misleading. Some of the features that each representation fails to predict in the other may be important parts of the computational solution that the two systems simply represent in incompatible ways. For example, in real data for complex tasks like language comprehension a few (important and interesting) dimensions are shared across individuals \citep{tuckute2025two}; however, the majority of the within-subjects explainable variance is not captured by these shared dimensions. Thus, a model that can predict the representations ``as well'' as the between-subjects noise ceiling might nevertheless miss some of the important computational features. The right panel of Fig. \ref{fig:rsa} shows a concrete example of how representation biases can contribute to this effect, where the multiple-easy model matches the representations of the multiple-hard models better than the multiple-hard models match each other---but the representations of the multiple-easy model \emph{do not} solve the hard tasks, whereas the less-aligned features of the multiple-hard models all consistently solve the same tasks.

Biased representations could also introduce challenges for analyses at lower levels. For example, in the experiments above, the units whose activity was most driven by the hard feature still tended to carry much \emph{stronger} signals about the easy feature (Fig. \ref{fig:biases_pca} right). Thus, these units might still get clustered with the units that only encode easy signals, or get removed from analyses due to their mixed selectivity \citep[cf.][]{rigotti2013importance}. If natural systems have similar unit-level biases, that would therefore similarly raise questions about whether the types of features we tend to identify a unit as encoding are necessarily the only important features about which that unit carries information.

Thus, representational biases could pose challenges for a range of common analyses. Of course, even with biases, there is still much that can be inferred from representational analyses; we will return to this point in the discussion. First, however, we will consider a more extreme case study.

\section{Homomorphic encryption: strongly dissociating computation from superficial patterns of representation}

In the experiments described above, the role that the representations played in the computations of the system was relatively straightforward, even where the representations were biased towards one feature. However, this does not have to be the case.
We illustrate this with a final conceptual case study of the possibility for strong dissociation between computation and patterns of representation: \emph{homomorphic} encryption \citep{gentry2009fully,van2010fully}. While the field of cryptography is largely focused on creating representations that preserve information yet are not easily decodable, in \emph{homomorphic} encryption schemes it is additionally possible to perform arbitrary computations (any algebraic circuit) over this information while it is encrypted. That is, at each step of such a computation, a new encrypted representation is produced that, when decrypted, yields the representation at that step of the original computation.

This example shows that it is not \emph{necessary} for a computational system to have any straightforward (e.g. linearly decodable) format of representation of the features that it uses in its computations. Systematic computations can be performed even over representations that are \emph{deliberately crafted} to thwart attempts to understand (decrypt) their content. 

As a special case, this example also illustrates that systematic compositional computations are possible without requiring representations that are straightforwardly (e.g. contatenatively) compositional in their original format. Encrypted representations are compositional only in the sense that ``with the right highly-nonlinear decoding scheme compositional representations can be extracted''---which is also true of some coding schemes typically interpreted as non-compositional, such as idiosyncratic representations of each input. That is, this case provides a clean example of a functionally-but-not-concatenatively compositional representation \citep[cf.][]{vangelder2014compositionality}. This raises questions about if and when it is feasible to rigorously confirm whether a system's computations are compositional from representational analyses. 

This example illustrates how opaque the relationship between the superficial format of representation and computation can be in the worst case.
However, computing using homomorphic encryption is generally too slow for many practical uses, which perhaps hints at one of the \emph{practical} constraints that pushes neural and machine learning systems toward simpler---and more aligned---representation formats; we will return to this question below.

\section{Discussion}

In this work, we have discussed some of the issues that arise in trying to interpret a system via its representations. We focused primarily on issues that arise due to biases that emerge in the representations learned by deep neural models trained to perform simple tasks---biases that favor certain features over others. We also highlighted the possibility for strong dissociation between representations and computations, as in the case of homomorphic encryption. Here, we discuss some natural questions that arise from these results, and how we see these issues relating to broader issues in the field.

\noindent
\textbf{Should we interpret these representational effects as biases?}

Above, we described the experimental results as representation ``biases.'' However, it is natural to ask whether this is an appropriate description. The patterns of bias are driven by real differences in the features being computed, and they have real downstream consequences (e.g. for the magnitude of representational perturbation needed to change the system's behavior). Thus, should we really interpret these effects as biases?

While we are sympathetic to this perspective, our central point is that these phenomena lead to biases in the inferences we draw about a system from using common representational analyses. Thus, whether these effects are interpreted as ``biases'' in the representations, they pose a challenge for interpreting frequently-used analyses. We therefore see them as a useful case study in the broader issues in how we analyze representations to understand a system.

\noindent
\textbf{Analyzing representations is still useful, even with biases---but complete understanding may be harder to achieve:}

It is important to note that these biases do not eliminate the possibility of achieving \emph{some} understanding from representational analysis. For example, the easy linear features that are readily discoverable in the models representations play an important, causal role in the model's behavior. The fact that the multi-task models and easy-task-only models appear very similar is \emph{precisely} due to the fact that (some of) their computations are similar. Moreover, biases in representation can lead to downstream biases in subsequent learning (\citealp{lampinen2024learned}; cf. \citealp{laamerad2024asymmetric})---thus, representation biases \emph{can} be important to how systems behave.

However, what the biases may limit is our ability to readily achieve \emph{complete} understanding. If representations are biased away from certain features, then \emph{some} important aspects of a system's behavior may be difficult to detect in its representations without \emph{a priori} knowledge of what features the system is representing. For example, the analyses used to quantify the role of different ways of representing the same nonlinear feature relied on knowing all the possible patterns of representing the feature in question,\footnote{Or rather, an augmented representation that is linearly transformable to any possible pattern of representation.} which is clearly impractical for natural systems in most cases.

These types of effects may result in a kind of ``streetlight effect'' where the field as a whole focuses on certain representational features that can be more reliably detected as strong signals in the neural activity, and ignores others that may nevertheless be important in the system's computations and behavior. This problem may be exacerbated if the easiest-to-detect signals also happen to be the simplest to interpret.

\noindent
\textbf{What are the potential solutions?}

It natural to ask whether alternative metrics for comparing systems could address some of these challenges. While considering a range of metrics is a generally good practice, it is not clear that there is a ready solution from metrics alone. First, there is often convergence of conclusions across different types of metrics \citep[e.g.][]{feather2025brain}---and theoretical results showing alignment \citep{williams2024equivalence}, even in some cases between seemingly-dissimilar metrics \citep{harvey2024representational}. Indeed, \citet{lampinen2024learned} found that the representation biases they observed produced similar biases in analysis results even when using less clearly variance-driven metrics, such cosine similarity. 
Moreover, even metrics that impose stronger constraints on the mapping between systems, such as soft matching distance \citep{khosla2024soft}, are vulnerable to similar issues---insofar as they still rely on variance-impacted measures like correlation at the unit level.
Of course, it would be possible to \emph{ignore} the variance carried by a signal when analyzing representations, but this would simply introduce another large bias---many components in the model's representations \emph{are} simply noise (in fact, all but two dimensions must ultimately be behavior-irrelevant in the simple model depicted above, due to its limited output rank). Thus, it will not necessarily be simple to address these issues by altering metrics alone.
Nevertheless, we believe that exploring whether modified metrics can help to ameliorate these issues is a valuable direction for future work. 

Second, as always causal interventions provide stronger confirmation of understanding. However, causal experiments do not fully address these issues on their own. 
Even in cases where causal interventions target unit- or population-level representations, these targets that we select for intervention are generally those that we have \emph{already} identified using tools such as regression, PCA, or clustering---and thus biases in these tools will mean that even if we verify the causal role of the representational features we identify, we still may not have identified the complete set of representational features that play a causal role in behavior. This problem is exacerbated by the possibility of causal preemption \citep[cf.][]{mueller2024missed}: strong interventions on a particular representational feature may effectively overwhelm the causal contribution of other features, leading us to believe we have identified the \emph{only} mechanism that mediates behavior, when in fact we have only identified one of multiple mechanisms that contribute. Verifying causality strengthens our conclusions about the contributions of a mechanism, but does not (on its own) guarantee that our understanding is complete.

Thus, one key to resolving the issues raised here is a strong focus on \emph{completely} characterizing behavior \citep[cf.][]{krakauer2017neuroscience} and the causal role that representations play in it. The feature biases we describe are defined precisely with respect to the behavioral outputs of the system. Thus, in the simple settings we consider, it would be easy to see that the strongly-represented components do not causally mediate some of the system's behaviors, and subsequently it might be possible to reverse-engineer how the system was representing them. Of course, for natural systems (with much richer input and output spaces), it may not be so easy to characterize behavior fully---nor to verify the causal role of a pattern of neural activity. Nevertheless, focusing on characterizing all the representational features that contribute to behavior will help to address these issues---especially if they are characterized across a broad range of varied settings (incorporating systematic manipulations) that may alter how computational mechanisms are engaged \citep[cf.][]{carvalho2025naturalistic}. 

\noindent
\textbf{If representations need not align, why \emph{do} we often observe convergence?}

Seeing that representations may be biased---or even encrypted---in ways that dissociate from their computational role makes it potentially more interesting that in practice we often observe representational alignment between systems \citep[e.g.][]{yamins2014performance,khaligh2014deep,nayebi2023mouse,schrimpf2021neural,sucholutsky2023getting,huh2024platonic}. One possible explanation is that the systems are primarily converging on the easiest features, which dominate the representations of all systems, regardless of whether those features are most important to the systems' computations.

However, a more optimistic possibility is that this convergence reflects real constraints on the types of representations that learning systems will tend to arrive at; for example, constraints on the solutions that are efficiently computable, on the solutions that satisfy a behavioral constraint \citep[e.g.][]{marchetti2024harmonics,piantadosi2024formalising}, or on the solutions that learning dynamics yield \citep[e.g.][]{saxe2022neural,smith2021origin,huh2021low}.
In this case, it may be productive to study more deeply the types of shared constraints on representation learning that drive alignment across systems (\citealp{cao2022multiple,cao2024explanatory}; cf. \citealp{huh2024platonic}).

\noindent
\textbf{The mutual benefits of knowledge sharing between neuroscience and machine learning}

At a higher level, we believe that these results illustrate the broader point that interactions between machine learning and neuroscience can be mutually beneficial. The ability to completely control the training of machine learning models, and to causally intervene on every component of them, allows us to explore issues like the reliability and faithfulness of analyses used in neuroscience \citep[cf.][]{jonas2017could}, and offers opportunities for exploring underlying assumptions. Reciprocally, neuroscience has a substantial history of analyzing complex systems---and identifying the challenges of bridging between levels of analysis---that is useful context for understanding the shared challenges in interpreting machine learning models \citep[cf.][]{he2024multilevel}.

\subsection*{Acknowledgments}

We thank Greta Tuckute, Nathaniel Daw, Lukas Muttenthaler, and Rapha{\"e}l Milli{\`e}re for helpful comments and suggestions on earlier drafts of this paper. We also thank Talia Konkle \& the Konklab, Daniel Yamins \& the NeuroAILab, and Nicolaus Kriegeskorte \& the Zuckerman Mind Brain Behavior Institute for stimulating discussions of these issues. We thank Alex Ku and Murray Shanahan for support.

\bibstyle{unsrtnat}
\bibliography{main}

\end{document}